# DEFLECTING CAVITY DYNAMICS FOR TIME-RESOLVED MACHINE STUDIES OF SXFEL USER FACILITY*

M. Song, H. Deng[†], B. Liu, D. Wang, SINAP, Shanghai, 201800, China

*Abstract*

Radio frequency deflectors are widely used for time-resolved electron beam energy, emittance and radiation profile measurements in modern free electron laser facilities. In this paper, we present the beam dynamics aspects of the deflecting cavity of SXFEL user facility, which is located at the exit of the undulator. With a targeted time resolution around 10 fs, it is expected to be an important tool for time-resolved commissioning and machine studies for SXFEL user facility.

## INTRODUCTION

Free electron laser (FEL) in the X-ray spectral region is a highly fruitful field ranging from ultra-fast scale probe to molecular biology, and from material science to medical science. Currently, the first X-ray FEL (8.8 nm) facility in China driven by 840 MeV LINAC is under construction at Shanghai, namely SXFEL test facility [1-2]. A soft X-ray user facility [3] has been proposed on the basis of SXFEL test facility. With a straightforward beam energy upgrade to 1.5 GeV, the FEL wavelength will extend to 2.0 nm and fully cover the water-window region [3]. In order to guarantee the FEL lasing performance at short wavelength, besides cascading HGHG [4], EEHG [5] and PEHG [6], a SASE [7] undulator line which consists of in-vacuum undulator and the insertion is also raised up.

For such an ultra-short bunch required for excellent FEL performance, one of the great challenges is the measurement and diagnosis with high temporal resolution. Up to now, many techniques have been developed, including zero RF phasing and streak camera. Transverse RF deflecting cavity is introduced to diagnose longitudinal profile of the electron bunch and FEL radiation, which is capable of resolving the temporal structure as short as sub-fs level under the circumstance of high deflecting voltage and frequency [8]. Since this method can effectively convert time-correlated longitudinal profile into the transverse profile, thus the bunch could be revealed and analysed in more detail. In terms of high efficiency and resolution, this technique would become key diagnostic system in the future. Therefore, a pair of X-band RF deflectors is planned at the exit of the undulator section of SXFEL user facility.

## DIAGNOSTIC BEAMLINE OPTIMIZATION

The preliminary designed deflector beamline of SXFEL user facility is shown in Fig. 1. Four quadruple magnets can be used for the beam optics optimization within the system, in which, one quadruple magnet downstream the RF deflecting cavity can be used for additional beam focusing in case of relative large beam size on the screen. The bending magnet located about 6 m downstream the undulator exit, is used for beam momentum and spread measurement. The transverse deflecting structure (TDS) installed in the beam line provide the performance to measure the longitudinal phase space, and thus the longitudinal bunch distribution.

The main goal of the whole TDS beamline is achieving high temporal resolution and allowing precious longitudinal measurement of ultra-short bunch. When the electron bunch passes through the RF deflecting cavity with the bunch centre at the zero-phase, the high frequency and time-resolved deflecting fields will kick the bunch and broaden its transverse size on the screen. According to the basics of the deflecting concept, the analytical formula of resolution can be deduced as follows [9]:

$$\Delta_s = \frac{c(E/e)}{\omega V_0} \frac{\sqrt{\varepsilon_x}}{\sqrt{\beta_d \sin \Delta\psi \cdot \gamma}} \quad (1)$$

On the basis of the formula, it is found that the resolution not only depends on the beam emittance and TWISS parameters but also influenced by the phase advance and deflecting force induced by the RF deflecting. Considering the time-resolved ability and the intrinsic beam profile, the beamline optimization could be achieved by adjusting four quadruple magnets gradient with ELEGANT [10] simulation under the condition that $\beta$ function at the screen ranging from 2 m to 4 m. The main parameters of TDS and optimized beamline are summarized in Table 1. It should be pointed out that the parameters are tentative and still need to be optimized.

Table 1: Main parameters of TDS and optimized beamline

| RF deflecting frequency | $f$ | 11.424 | $GHz$ |
|---|---|---|---|
| RF deflector voltage | $V_0$ | 10 | $MV$ |
| RF deflector length | $L$ | 1 | $m$ |
| RF deflector number |  | 2 |  |
| Nominal beam size | $\sigma_{x0}$ | 31.5 | $\mu m$ |
| Beam size with TDS on | $\sigma_x$ | 971.6 | $\mu m$ |
| Beta at TDS | $\beta_d$ | 8.8 | $m$ |
| Beta at screen | $\beta_s$ | 2.9 | $m$ |
| Phase advance | $\Delta\varphi$ | 79 | $degree$ |
| Normalized emittance | $\varepsilon_n$ | 1.015 | $\mu m$ |
| RMS bunch length | $\sigma_z$ | 55.797 | $\mu m$ |

___________________

*Work supported by Natural Science Foundation of China (11475250 and 11322550) and Ten Thousand Talent Program.
[†] denghaixiao@sinap.ac.cn

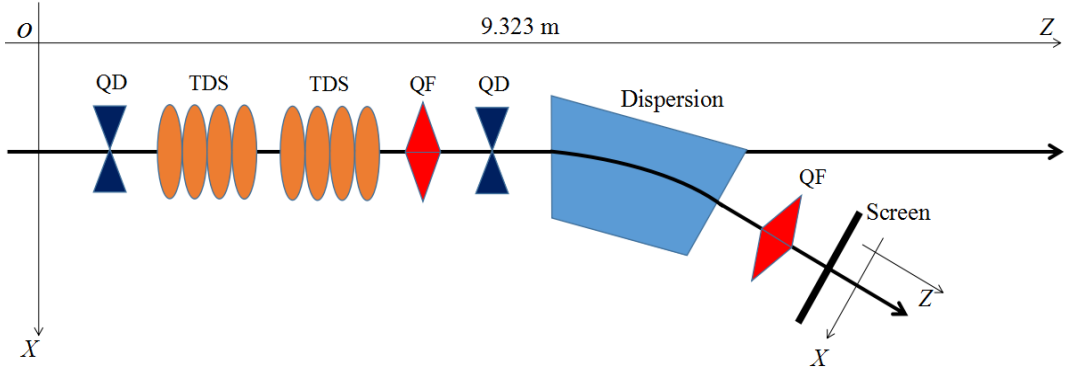

Figure 1: Layout of the TDS beamline downstream the undulator section

According to the main parameter from Table 1 and analytical formula, the best time resolution can be achieved is approximately 6.6 fs with two RF deflectors of this type, which is pretty enough to measure 180 fs electron bunch in SXFEL user facility.

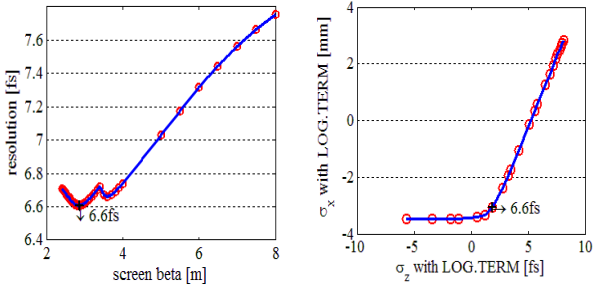

Figure 2: Resolution optimization & simulation results

Figure 2 presents the resolution results from different method, the left shows the resolution distribution versus screen beta through adjusting four quadruple with regarding screen beta as target, then the computed phase advance and beta at deflector substituted into analytical formula will generate the resolution profile, it is obvious to find the minimum resolution approaches nearly 6.6 fs. The right shows the case on the basis of the optimized lattice from former method, in this sense, RMS bunch length was increased from almost zero to 1 mm. It is deserved noting that the horizontal RMS beam sizes on the screen approximately keep constant at first, while with the growth of bunch length, the beam size will experience a linear relationship with bunch length. According to the resolution definition and Fig. 2 (b), the transverse beam size at the screen is square root of two times of initial size when the RMS bunch length reaches about 6.6 fs, which also corresponding to the best resolution value and verifying the correction of previous lattice optimization.

## LONGITUDINAL PHASE SPACE & BUNCH PROFILE MEASUREMENT

The bunch length and longitudinal profile measurements performed by the RF deflecting cavities installed at the SXFEL user facility diagnostic beamline need horizontal focus on the downstream screen. As a result, the beam longitudinal profile can be reconstructed by the horizontal image on the screen.

*TDS calibration*

The horizontal position of bunch on the screen at the end of beamline wholly depends on the RF deflecting phase at which bunch experiences. In order to reconstruct longitudinal profile one needs to know the deflecting parameter which could be measured in an experiment, here also could be obtained through ELEGANT simulation [10]. It is assumed that TDS RF phase is earlier or later than the zero-phase at which bunch propagates. Thus the electron bunch at the previous position will experience an off-axis shift due to transverse kick. As a result, considering shifted length and deriving it in terms of phase delay, the deflecting parameter can be written as follows:

$$D = K * 360° * f / (\beta * c) \qquad (2)$$

Here K is slope of the linear line with f the deflecting frequency. For the SXFEL user facility diagnostic beamline, the simulation result of TDS calibration is shown in Fig. 3.

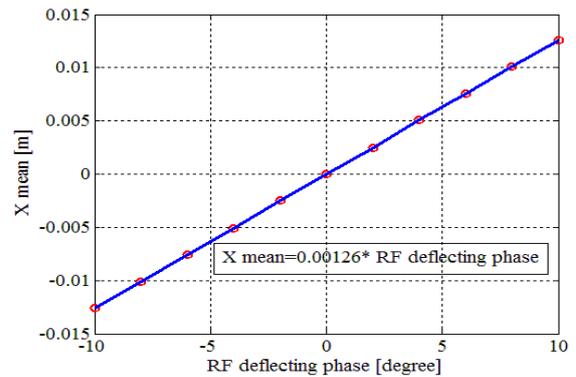

Figure 3: TDS calibration at the observation screen at the end of beamline. The horizontal beam centre shows a dependence on RF deflecting phase with red circle points, the blue line represents the linear fit.

On the basis of the eleven groups of simulation results one can obtain the slope of linear fit is 0.00126 m/degree, thus for user facility diagnosis parameter with a frequency of $f$=11.424 $GHz$ and relativistic velocity equal to one. Equation (2) can calculate the deflecting parameter as 17.273.

*Longitudinal phase space reconstructed*

In future measurements of longitudinal phase space at the SXFEL user facility could be done with the combination of RF deflecting cavity and dipole magnet as well as to achieve higher resolution. As aforementioned, the diagnostic beamline should be optimized along with deflecting parameter measured to allow for such measurement. In addition, a more detailed schematic layout of the measurement items related to longitudinal phase space has shown in Fig. 1.

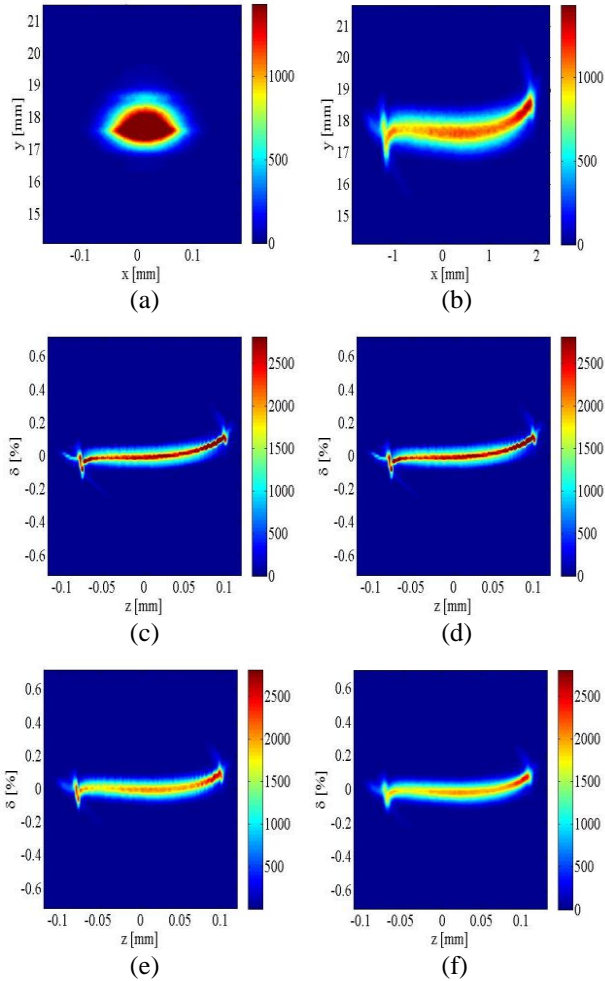

Figure 4: Simulated screen beam images with TDS off (a) and TDS on (b). Simulated longitudinal phase space along the beamline: (c) the original phase space, (d) in front of the RF deflecting cavity, (e) in front of the dipole magnet, (f) reconstructed phase space at the screen.

Simulated images on the observation screen are presented in Fig. 4. The picture in Fig. 4 (a) shows the beam image on the observation screen when the RF deflecting cavity is switched off. While the Fig. 4 (b) presents beam image with TDS switched on and this will give a longitudinal resolution of 6.6 fs. It is obvious to find that two RF deflecting cavities operated in X-band give the electron bunch a strong horizontal kick. Simulation consequences of longitudinal phase space are presented in Fig. 4 with electron bunch charge 500 pC. The plot (c) shows the original electron bunch longitudinal phase space. Moreover, the plot (d) expresses the longitudinal phase space in front of RF deflecting cavity. Furthermore, the plot (e) presents the longitudinal phase space downstream the TDS while upstream the dispersion section. In the final, the plot (f) represents the reconstructed longitudinal phase space located at the observation screen, which derived from Fig. 4 (b).

From the longitudinal phase space evolution along the diagnostic beamline, one can find that the reconstructed one shown in Fig. 4 (f) becomes wider and has a slightly different shape from input electron bunch profile. This effect can be explained by the additional increased TDS energy spread.

*Longitudinal bunch profile reconstructed*

The reconstructed longitudinal profile is shown in Fig.5 resulted from Fig. 4 (f), which scaled screen image size with deflecting parameter of 17.273. The longitudinal profile obtained through ELEGANT [10] particle tracking are also shown. The Fig. 5 shows a good agreement between reconstructed profile and initial one.

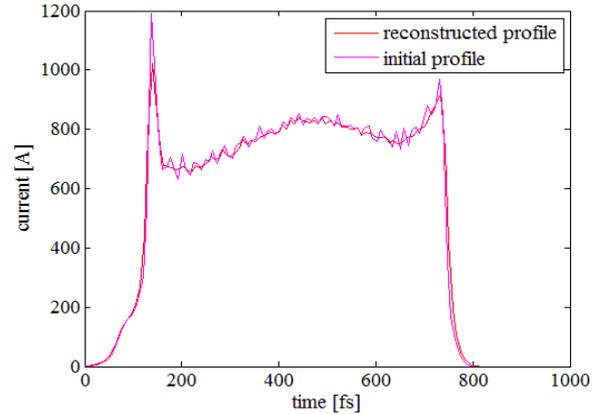

Figure 5: Comparison between the tracked current profile (pink curve) and the reconstructed profile (red curve).

## FEL PROFILE RECONSTRUCTION

The direct application of RF deflecting cavity is the longitudinal phase space observation of the electron beam. As demonstrated at LCLS [8], under the verification of simulated results of reconstructed longitudinal bunch profile compared with given distribution, the SXFEL diagnostic beamline has the capability to measure the X-ray pulse temporal distribution.

As presented in Fig. 6, the longitudinal profile of the X-ray FEL pulse are simulated and reconstructed, which illustrates that the electron beam experience a two-stage HGHG lasing, by comparing the longitudinal phase with FEL lasing-off, the FEL pulse duration and shape can be retrieved as shown in Fig. 6 (b).On the basis of the temporal profile, on can find an approximately 1.5 GW radiation power generated at the first stage HGHG along with a nearly 400 MW X-ray power produced at the second stage HGHG.

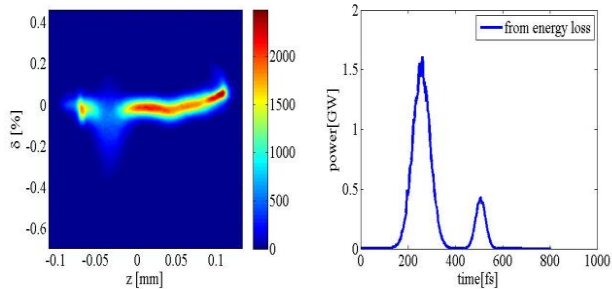

Figure 6: FEL temporal profile reconstruction. The electron bunch charge is 500 pC with electron energy of 1.5 GeV. The simulated results are shown in: (a) Reconstructed FEL lasing downstream the TDS, (b) Reconstructed X-ray pulse duration and shape.

## CONCLUSION&OUTLOOK

The schematic layout of TDS beamline for SXFEL user facility was described and optimized in this paper. With the installation of X-band RF deflecting cavities, measurements of the longitudinal electron beam phase space and the X-ray FEL pulse profile can be carried out. Simulation results confirm that the beamline allows one to measure the electron bunch with a time resolution of 6.6 fs, which can be further enhanced by the transverse gradient undulator compensation [11]. In addition, besides the capabilities shown in this paper, other potential applications, for instance, studies of micro-bunching instability and slice energy spread can be carried out in the future.

## REFERENCES

[1] Soft X-ray FEL Concept Design Report, Shanghai, 2015.

[2] M. Song *et al.*, Nucl. Instr. and Meth. A822, 71 (2016).

[3] M. Song*et al*.,in: Proceedings of IPAC16, Busan, Korea, 2016.

[4] L.H. Yu, Phys. Rev. A 44, 5178(1991).

[5] G. Stupakov, Phys. Rev. Lett. 102, 074801 (2009).

[6] H. Deng, C. Feng, Phys. Rev. Lett. 111, 084801(2013).

[7] A. Kondratenko, E. Saldin, Part Accel. 10, 207 (1980).

[8] C. Behrens*et al*., Nature Communication 5, 3762 (2014).

[9] R. Akre *et al*., SLAC-PUB-8864, June, 2001.

[10] M. Borland *et al*., Elegant: A flexible SDDS-compliant code for accelerator simulation, Argonne National Lab, IL, US, 2000 (unpublished).

[11] G. Wang *et al*., Time-resolved electron beam diagnostics with sub-femtosecond resolution, arXiv: 1510.06111.